\documentclass[fleqn,usenatbib]{mnras}

\def\lesssim{\mathrel{\hbox{\rlap{\hbox{\lower4pt\hbox{$\sim$}}}\hbox{$<$}}}}
\def\gtrsim{\mathrel{\hbox{\rlap{\hbox{\lower4pt\hbox{$\sim$}}}\hbox{$>$}}}}

\newcommand{\ips}{\ensuremath{i_{\rm P1}}}

\newcommand{\wps}{\ensuremath{w_{\rm P1}}}
\newcommand{\grizy}{\ensuremath{grizy_{\rm P1}}}

\newcommand{\degree}{\mbox{$^\circ$}}

\newcommand{\msol}{\mbox{M$_{\odot}$}}

\newcommand{\kms}{\mbox{$\rm{km}\,s^{-1}$}}

\def\GWsource{GW190425}
\def\UTGW{2019-04-25 08:18:05 UT}  
\def\MJDGW{58598.34589}       
\def\alertMJDGW{58598.4118}          
\def\GWdist{$159^{+69}_{-72}$\,Mpc}     

\def\MJDPSo{58598.40265}           
\def\MJDdiffoPS{1.36\,hrs}         
\def\MJDdiffgal{6.0\,hrs}         
\def\MJDdiffoATLAS{0.8\,hrs}         
\def\initialMap{\texttt{bayestar.fits}} 
\def\FRBsource{FRB\,190425}
\def\MJDFRB{58598.44899}   
\def\UTFRB{2019 Apr 25 10:46:33}    
\def\FRBredshiftlim{$z<0.0394$}
\def\FRBdistancelim{$D_{\rm L}<179$\,Mpc}
\def\HostGal{UGC\,10667}     
\def\HOSTCoordRA{255.662479}
\def\HOSTCoordDEC{21.576746}

\def\Hostz{$0.031224 \pm 0.000011$}     
\def\HostDist{$141\pm10$\,Mpc}     
\def\EBsubV{0.066}                  
\def\Ag{0.247}                  
\def\Ar{0.171}                  
\def\Ai{0.127}                  

\usepackage{newtxtext,newtxmath}
\usepackage{xcolor}

\usepackage[T1]{fontenc}
\usepackage{orcidlink}
\DeclareRobustCommand{\VAN}[3]{#2}
\let\VANthebibliography\thebibliography
\def\thebibliography{\DeclareRobustCommand{\VAN}[3]{##3}\VANthebibliography}

\usepackage{graphicx}	
\usepackage{amsmath}	

\title[Optical constraints on GW190425 and FRB190425]{
GW190425: Pan-STARRS and ATLAS coverage of the skymap and limits on optical emission associated with FRB190425}

\author[Smartt et al.]{
S. J. Smartt$^{1,2}$\thanks{E-mail:stephen.smartt@physics.ox.ac.uk}\orcidlink{0000-0002-8229-1731}, 
M. Nicholl$^{2}$\orcidlink{0000-0002-2555-3192},   
S. Srivastav$^{2}$\orcidlink{0000-0003-4524-6883},   
M. E. Huber$^{3}$\orcidlink{0000-0003-1059-9603},    
K. C. Chambers$^{3}$\orcidlink{0000-0001-6965-7789},   
K. W. Smith$^{2}$\orcidlink{0000-0001-9535-3199},      
\newauthor{
D. R. Young$^{2}$\orcidlink{0000-0002-1229-2499},     
M. D. Fulton$^{2}$\orcidlink{0000-0003-1916-0664},    
J. L. Tonry$^{3}$\orcidlink{0000-0003-2858-9657}, 
C. W. Stubbs$^4$\orcidlink{0000-0003-0347-1724}, 
L. Denneau$^{3}$, 
A. J. Cooper$^{1}$\orcidlink{0000-0002-4033-3139}, 
A.  Aamer$^{2}$,    
}
\newauthor{
J. P. Anderson$^{5,6}$\orcidlink{0000-0003-0227-3451}, 
A. Andersson$^1$\orcidlink{0000-0003-2734-1895}, 
J. Bulger$^{3}$, 
T.-W Chen$^{7}$\orcidlink{0000-0002-1066-6098}, 
P. Clark$^8$\orcidlink{0000-0002-6576-7400}, 
T. de Boer$^{3}$,    
H. Gao$^{3}$\orcidlink{0000-0003-1015-5367}, 
}
\newauthor{
J. H. Gillanders$^{1}$\orcidlink{0000-0002-8094-6108}, 
A. Lawrence$^{3}$\orcidlink{0000-0002-4474-4963}, 
C. C. Lin$^{3}$\orcidlink{0000-0002-7272-5129}, 
T. B. Lowe$^{3}$, 
E. A. Magnier$^{3}$\orcidlink{0000-0002-7965-2815}, 
P. Minguez$^{3}$,
T. Moore$^{2}$\orcidlink{0000-0001-8385-3727}, 
}
\newauthor{
A. Rest$^{9,10}$\orcidlink{0000-0002-4410-5387}, 
L. Shingles$^{11}$\orcidlink{0000-0002-5738-1612}, 
R. Siverd$^{3}$\orcidlink{ 0000-0001-5016-3359}, 
I. A. Smith$^{12}$\orcidlink{0000-0001-8605-5608}, 
B. Stalder$^{13}$\orcidlink{0000-0003-0973-4900}, 
H. F. Stevance$^{1}$ 
R. Wainscoat$^{3}$\orcidlink{0000-0002-1341-0952}, 
}
\newauthor{
R. Williams$^{14}$ 
}
\\
$^{1}$Astrophysics sub-Department, Department of Physics, University of Oxford, Keble Road, Oxford, OX1 3RH, UK\\
$^{2}$Astrophysics Research Centre, School of Mathematics and Physics, Queen's University Belfast, BT7 1NN, UK\\
$^{3}$Institute for Astronomy, University of Hawai'i, 2680 Woodlawn Drive, Honolulu, HI 96822, USA\\
$^4$Department of Physics and Department of Astronomy, Harvard University, Cambridge, MA 02138, USA\\
$^5$European Southern Observatory, Alonso de C\'{o}rdova 3107, Casilla 19, Santiago, Chile\\
$^6$Millennium Institute of Astrophysics (MAS), Nuncio Monse\~{n}or S\'{o}tero Sanz 100, Off. 104, Providencia, Santiago, Chile\\
$^7$ Graduate Institute of Astronomy, National Central University, 300 Jhongda Road, 32001 Jhongli, Taiwan\\
$^{8}$Institute of Cosmology and Gravitation, University of Portsmouth PO1 3FX, Portsmouth, UK\\
$^{9}$Space Telescope Science Institute, 3700 San Martin Drive, Baltimore, MD 21218, USA\\
$^{10}$Department of Physics and Astronomy, Johns Hopkins University, Baltimore, MD 21218, USA\\
$^{11}$GSI Helmholtzzentrum f\"{u}r Schwerionenforschung, Planckstrasse 1, D-64291 Darmstadt, Germany\\
$^{12}$ Institute for Astronomy, University of Hawai'i, 
34 Ohia Ku St., Pukalani, HI 96768-8288, USA\\
$^{13}$Vera C. Rubin Observatory Project Office, 950 N Cherry Ave, Tucson, AZ 95719, USA\\
$^{14}$ Institute for Astronomy, University of Edinburgh, Royal Observatory, Blackford Hill, EH9 3HJ, UK\\
}

\date{Submitted 20th September 2023}

\pubyear{2023}

\begin{document}
\label{firstpage}
\pagerange{\pageref{firstpage}--\pageref{lastpage}}
\maketitle

\begin{abstract}
\GWsource\ is the second of only two binary neutron star (BNS) merger events to be significantly detected by the LIGO-Virgo-Kagra gravitational wave detectors. With  a detection only in LIGO Livingston, the skymap containing the source was large and no plausible electromagnetic counterpart was found in real time searching in 2019. Here we summarise our ATLAS and  Pan-STARRS wide-field optical coverage of the skymap beginning within 1 hour and 3 hours respectively of the GW190425 merger time. More recently, a potential coincidence between  \GWsource\ and a fast radio burst \FRBsource\ has been suggested, given their spatial and temporal coincidence. The smaller sky localisation area of \FRBsource\ and its dispersion measure have led to the identification of a likely host galaxy, 
\HostGal\ at a distance of \HostDist.  Our optical imaging covered the galaxy \MJDdiffgal\ after \GWsource\ was detected and 3.5\,hrs after the \FRBsource. No optical emission was detected and further imaging at +1.2 and +13.2 days also revealed no emission. If the \FRBsource\ and \GWsource\ association were real, we highlight our limits on kilonova emission from a BNS merger in \HostGal.  The model for producing \FRBsource\ from a BNS merger involves a supramassive magnetised neutron star  spinning down by dipole emission on the timescale of hours. We show that magnetar enhanced kilonova emission is ruled out by optical upper limits. 
The lack of detected optical emission from a kilonova in \HostGal\ disfavours, but does not disprove, the FRB-GW link for this source. 

\end{abstract}



\section{Introduction}

The historic gravitational wave event GW170817 resulting from a binary neutron star (BNS) merger \citep{2017PhRvL.119p1101A}  produced a short gamma-ray burst \citep[GRB170817A;][]{2017ApJ...848L..13A} and a rapidly evolving optical and infrared transient \citep[AT2017gfo;][]{MMApaper2017}. The  relatively small sky localisation map (31 square degrees), 
inferred from the strong signals in the two LIGO detectors and an upper limit in Virgo 
\citep{2017PhRvL.119p1101A},  allowed the rapid identification of an optical 
counterpart.  This was achieved 11 hours after the BNS merger, 
during the first night of observing the sky map from Chile 
\citep{Arcavi2017,Coulter2017,Lipunov17,SoaresSantos2017,Tanvir2017,Valenti2017}.  Global monitoring followed, with the spectra from Chilean and South African observatories showing an unprecedented evolution within the firts 24hrs and confirming that this was the  signature of a unique transient with no known counterpart \citep{Chornock2017,McCully2017,Nicholl2017,2017Natur.551...67P,Shappee2017,Smartt2017}. 
The lightcurve monitoring showed AT2017gfo faded rapidly with the flux emission shifting to near infra-red and possibly even beyond
\citep{Andreoni2017,Cowperthwaite2017,Drout2017,Evans2017,Kasliwal2017,Kilpatrick2017,Tanvir2017,Troja2017,Utsumi2017}.
The transient was detected in the x-ray and radio a few days after the merger \citep{Alexander2017,Haggard2017,Hallinan2017,Margutti2017,Troja2017}, 
providing constraints on the jet physics giving rise to the short GRB.

GW170817 was discovered towards the end of the LIGO-Virgo collaboration's second observing run (O2) and at the time of writing (3 months into O4) only one further BNS merger has been significantly  detected.
GW190425 was observed in only one LIGO detector \citep{GW190425LVC} close to the 
start of the 3rd Observing run (O3). With a signal only from Livingston, 
the sky 
localisation map was very large and half of the high probability region was in the 
day time sky. No electromagnetic counterpart was discovered at the time at any wavelength 
\citep{2019ApJ...885L..19C,2019GCN.24210....1S,2021A&A...650A.131B,2021ApJ...912..128P}
which was not a major surprise given the large sky map, the inferred distance $D_L$=\GWdist\ 
\citep[from the GW analysis of][]{GW190425LVC}
and solar conjunction. 

Fast radio bursts (FRBs) are extra-galactic millisecond-duration bursts of unknown origin. The large all-sky rate of FRBs appears inconsistent with a single compact object merger origin of all sources \citep{Ravi2019}. However, some models suggest mergers may be responsible for a subset of FRBs, powered by either pre-merger magnetic interaction (e.g. \citealt{Totani2013}) or the merger remnant (e.g. \citealt{2014A&A...562A.137F}). These models can be effectively tested by performing prompt FRB searches on localization regions of gravitational wave events and short gamma-ray bursts, or through post-FRB follow-up observations of nearby sources in search of a kilonova or radio afterglow \citep{Cooper2023}.

A search for spatial and temporal coincidences of gravitational wave events and FRBs by \cite{FRB190425} proposed a link between 
\FRBsource\ and GW190425. They searched for
GW-FRB coincidences with the Canadian Hydrogen Intensity Mapping Experiment FRB catalogue (CHIME/FRB) 
\citep{2021ApJS..257...59C}
and the O3 GWTC-2 catalogue
\citep{2021PhRvX..11b1053A}. 
They used a time window
of 26hrs, from 2hrs before the GW event and
up to 24hrs after. The CHIME sky localisation 
region is typically of order arcminutes in diameter, much smaller than  that of GW sources (tens to hundreds of square degrees depending on the number
of detectors retrieving a signal). 

\FRBsource\ was detected at \MJDFRB\ 
\citep[\UTFRB UT;][]{2021ApJS..257...59C}, which was 
2.5\,hrs after \GWsource\ \citep{FRB190425}. \cite{FRB190425} report that its estimated sky position of 
RA$=255.72\pm0.14^{\circ}$
DEC$=21.52\pm0.18^{\circ}$
places it within the 66.7 per cent probability 
countour of the final, most reliable sky map from 
GWTC-2 
\citep{2021PhRvX..11b1053A}. In addition, the dispersion measure (DM)
of \FRBsource\ provides an upper limit to the
redshift of \FRBredshiftlim\ corresponding 
to a luminosity distance\footnote{We assume a standard flat cosmology with $H_0 = 68$\kms\,Mpc$^{-1}$  from the 
\cite{Planck2016} as adopted in \cite{GW190425LVC}.}
of \FRBdistancelim. The FRB and GW signals were therefore coincident in their sky positions, 
distances and time (given the definition of coincidences described). \cite{FRB190425}
propose that there is only one catalogued
galaxy (in NED, the NASA/IPAC Extragalactic Database)
within the 
CHIME error ellipse that has a measured spectroscopic redshift placing it within the 
upper limit measured from the DM
of \FRBredshiftlim. This is \HostGal,  at a 
redshift of \Hostz\ \citep[from SDSS Data Release 13]{2017ApJS..233...25A}. We adopt a foreground 
extinction toward \HostGal\ of 
$E(B-V)$=\EBsubV, 
corresponding to 
$A_g$=\Ag, 
$A_r$=\Ar, 
$A_i$=\Ai, 
\citep{2011ApJ...737..103S}.

\cite{FRB190425host} further investigated the 
plausibility of \HostGal\  being the host of  
FRB190425. Both \cite{FRB190425host} and \cite{FRB190425} 
selected \HostGal\  as the most probable host due to its 
location within the 68\% localization uncertainty 
of \FRBsource\ (from CHIME's ellipse, of dimensions $0.1 \times 0.2$\,deg).  
The redshift
corresponds to  a Hubble flow distance of $D_L=$\HostDist\ (corrected for Virgo infall, from NED),  consistent with the LIGO-Virgo distance constraint 
 $D_L$=\GWdist\ \citep{GW190425LVC} and the 
upper limit to the FRB DM   \FRBdistancelim. 
\cite{FRB190425host} find that UGC10667 is a spiral galaxy with a modest star formation rate and luminosity dominated by an old stellar population. 
They also searched for  transient radio  emission in and around the galaxy at 2.5\,yrs post-burst that could be associated with either the 
\FRBsource\ or \GWsource\ and found no convincing radio transient emission in VLA 6\,Ghz data
taken in September and October 2021. 

\cite{FRB190425} and \cite{FRB190425host} highlight that \FRBsource\ had a number of 
notable properties; one of lowest DM non-repeating events in the CHIME FRB Catalogue 1, 
a high flux for those with low DMs, a short pulse duration and a flat spectrum. 
All of this led \cite{FRB190425} to suggest that the co-production of  
\GWsource\ and \FRBsource\ could be explained by the theory of \cite{2014ApJ...780L..21Z}. 
In this scenario the BNS merger produces a supramassive neutron star, which is highly 
magnetised. The compact object has a short rotation period and loses angular momentum as it 
spins down and collapses to a black hole. The FRB is created as the magentosphere is 
ejected \citep{2014A&A...562A.137F}, through the so-called ``blitzar'' mechansim. 
The supramassive neutron star must survive for 2.5hrs, the time between merger and 
the FRB. While the data and theory of association are intriguing, \cite{2023arXiv230600948B} have cautioned against assuming physical association from analysis of the GW signal and constraints on the ejecta mass for the 400MHz radio signal to propagate. 

In this paper we present a summary of the ATLAS and Pan-STARRS wide-field optical coverage  of the GW190425 skymap starting\
\MJDdiffoATLAS\ and  \MJDdiffoPS\ respectively  
after the BNS merger. We present images of the proposed most likely host of \FRBsource\ (\HostGal) taken over the first few nights, finding no optical transient emission. We also report publicly available Zwicky Transient Facility data \citep{2019PASP..131a8002B} of the host. 
We 
discuss the 
plausibility of the GW-FRB link assuming that the host galaxy is indeed \HostGal\ and the BNS produced a kilonova through mass ejection. 

\begin{figure*}
	\includegraphics[width=18cm]{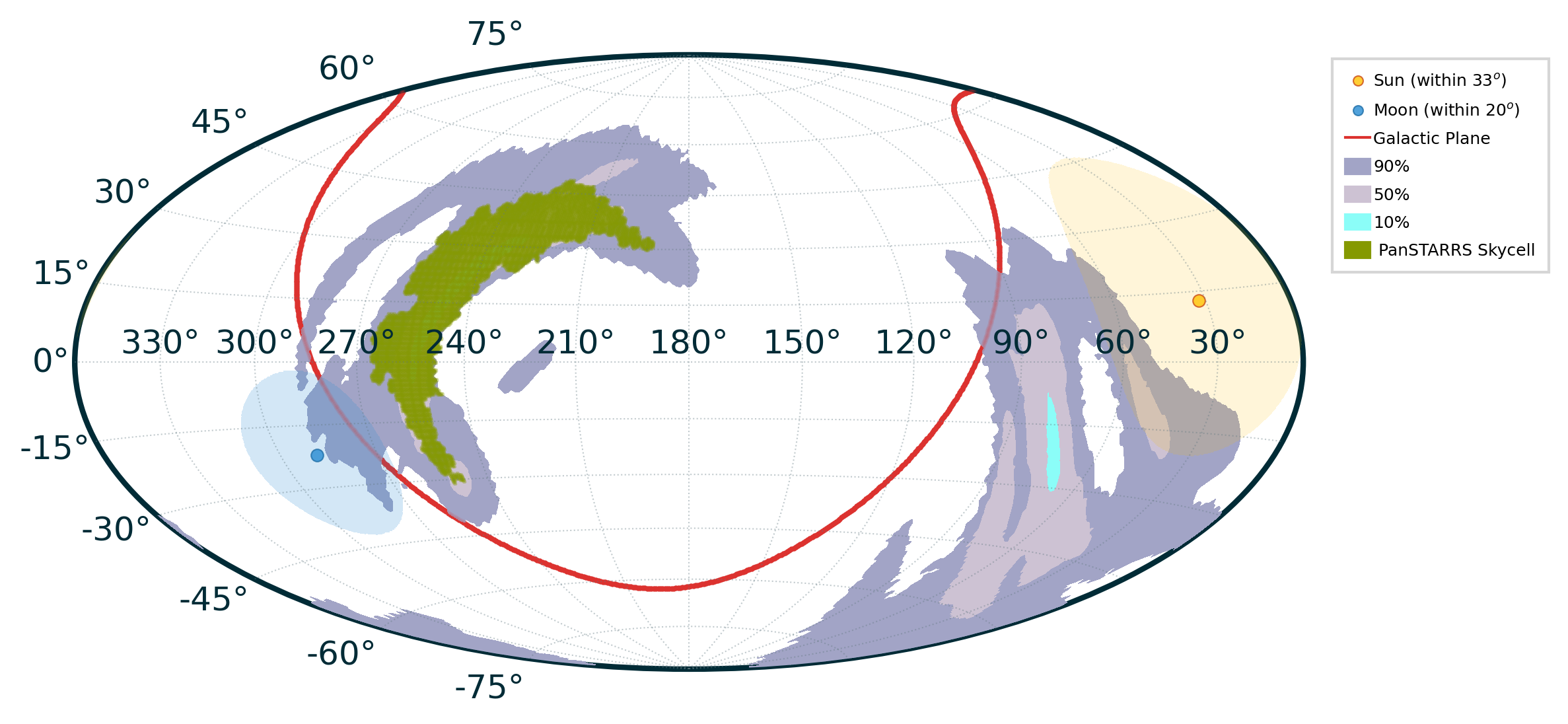}
     \caption{The Pan-STARRS1 coverage over the first three nights of observing starting \MJDdiffoPS\,hrs after the detection of \GWsource. We used the \texttt{bayestar.fits} skymap to define our pointings at the time, 
but the map above shows the final GWTC-2 skymap as released on GraceDB. The cumulative probability covered at after one night was 23.8 per cent, which was incremented to 24.9 per cent after three nights. 
    }
    \label{fig:PS1coverage}
\end{figure*}

\begin{figure*}
	\includegraphics[width=18cm]{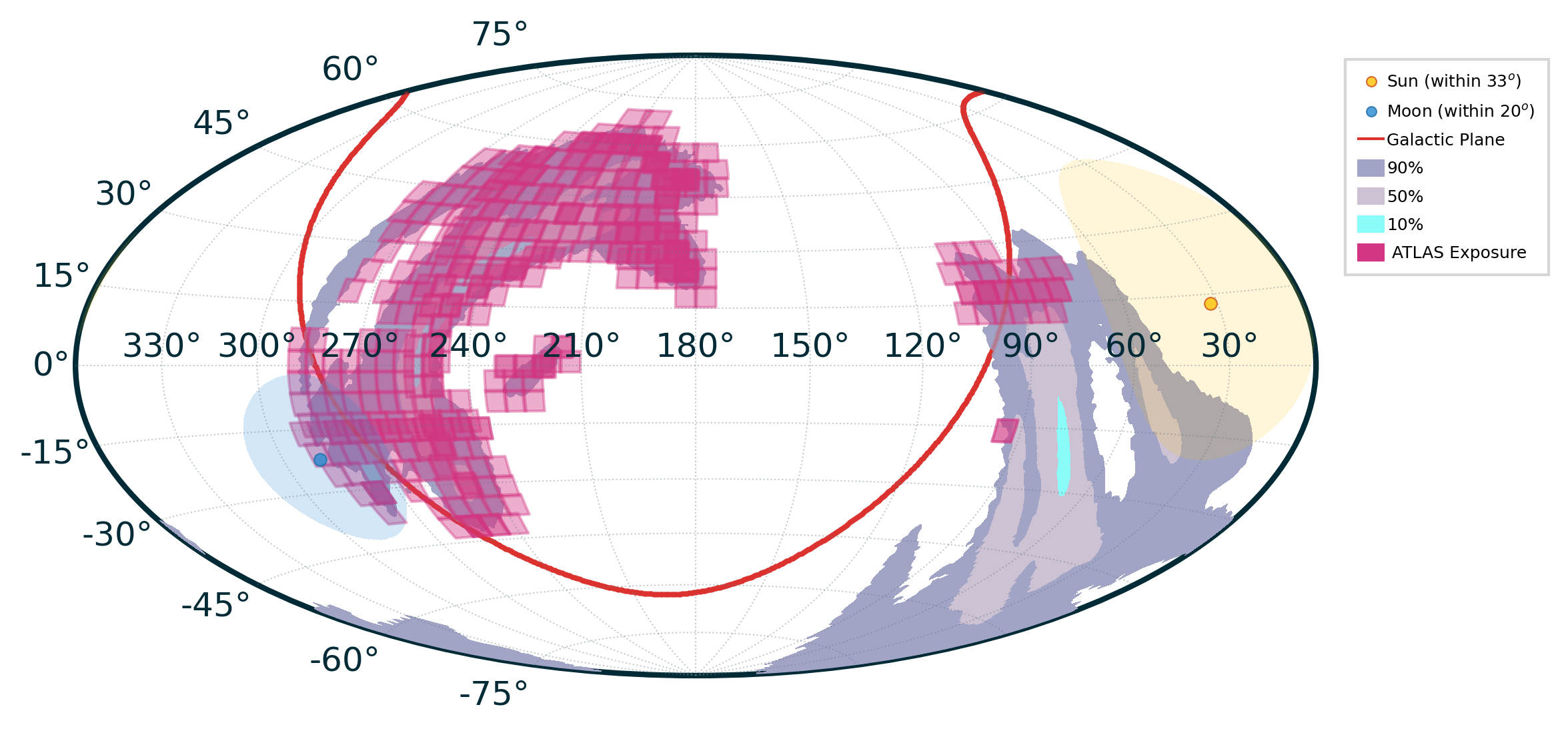}
     \caption{The ATLAS coverage over the first three nights of observing starting 
     \MJDdiffoATLAS\,hrs after the detection of \GWsource. We used the \texttt{bayestar.fits} skymap to define our pointings at the time, 
    but the map above shows the final GWTC-2 skymap as released on GraceDB.  The cumulative probability covered at this stage was 41.6 per cent. 
    }
    \label{fig:ATLAScoverage}
\end{figure*}

\begin{figure}
	\includegraphics[width=\columnwidth]{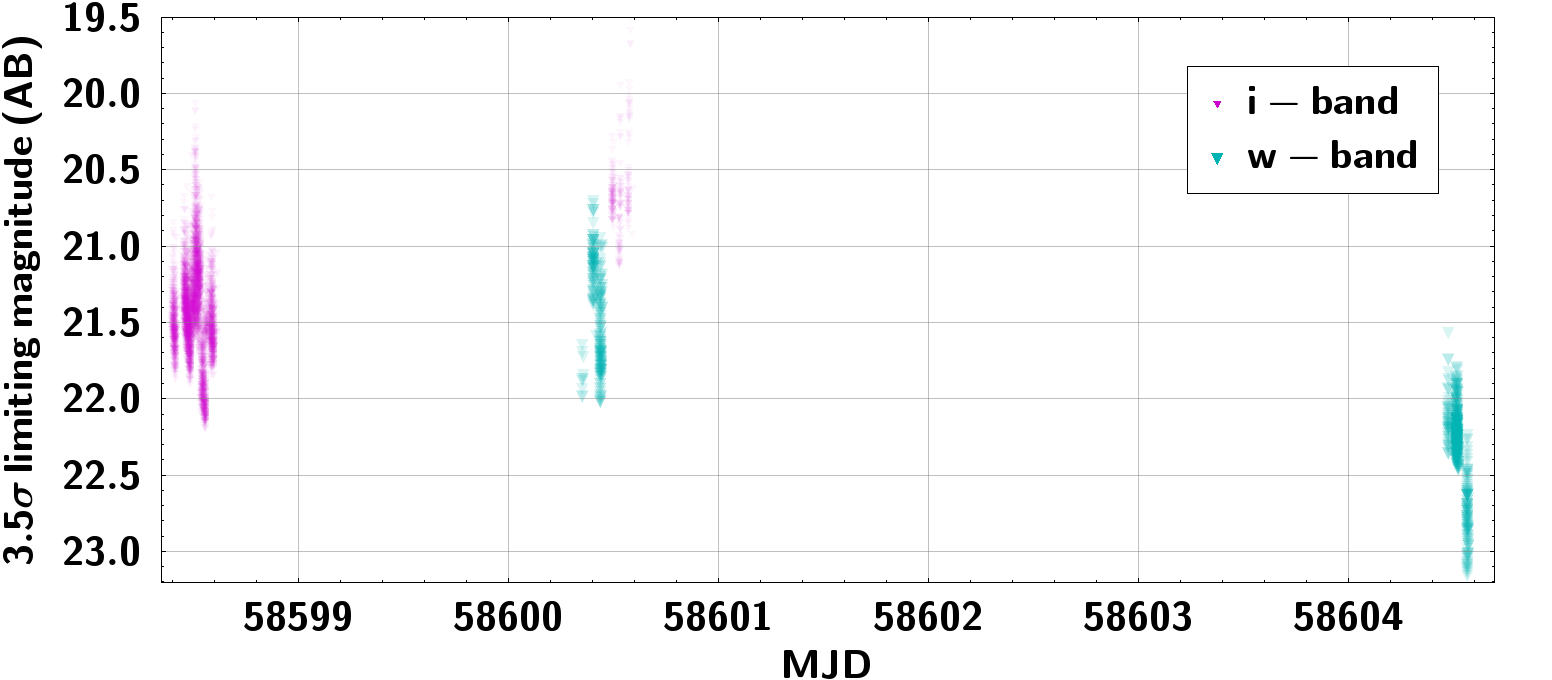}
	\includegraphics[width=\columnwidth]{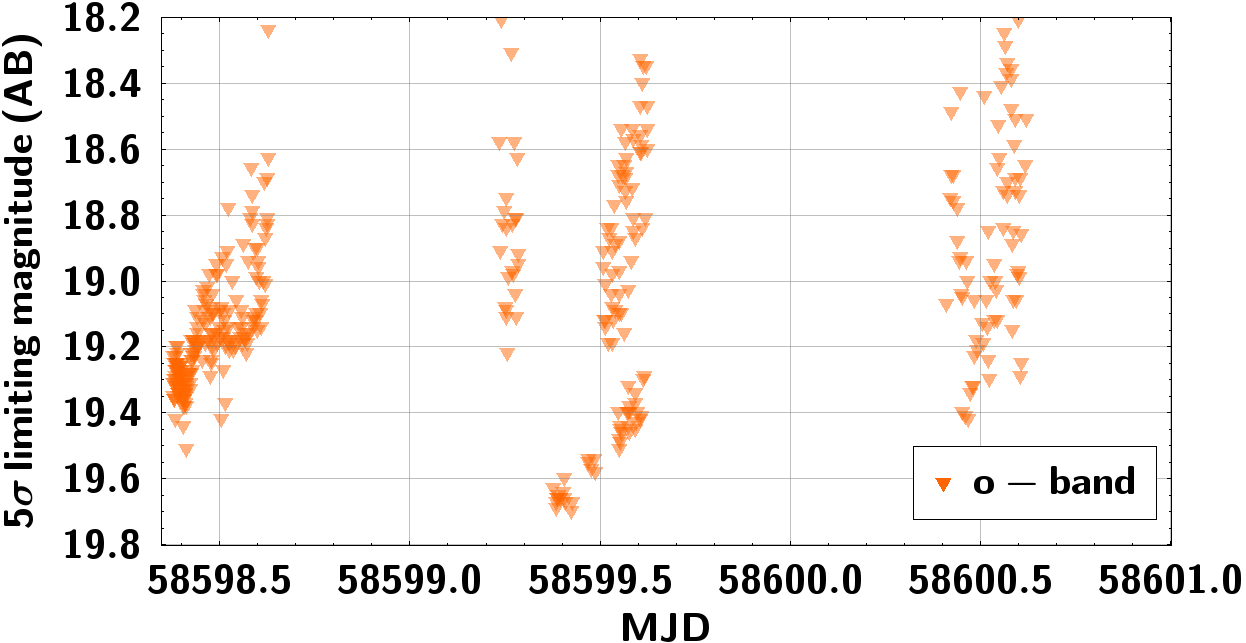}
    \caption{Illustration of the limiting magnitudes of Pan-STARRS1 (top) and ATLAS imaging (bottom) of images that fell within the 90 per cent probability of the sky map \texttt{gw190425z$\_$skymap.multiorder.fits}. The left hand side of the axis is set to the GW detection time \MJDGW. Each Pan-STARRS1 point represents a single skycell of the processed GPC1 frame with exposure times of 45 sec, and the ATLAS points represent a single full-frame ATLAS camera footprint of 30 seconds. 
    }
    \label{fig:limits}
\end{figure}

\begin{figure*}
	\includegraphics[width=\textwidth]{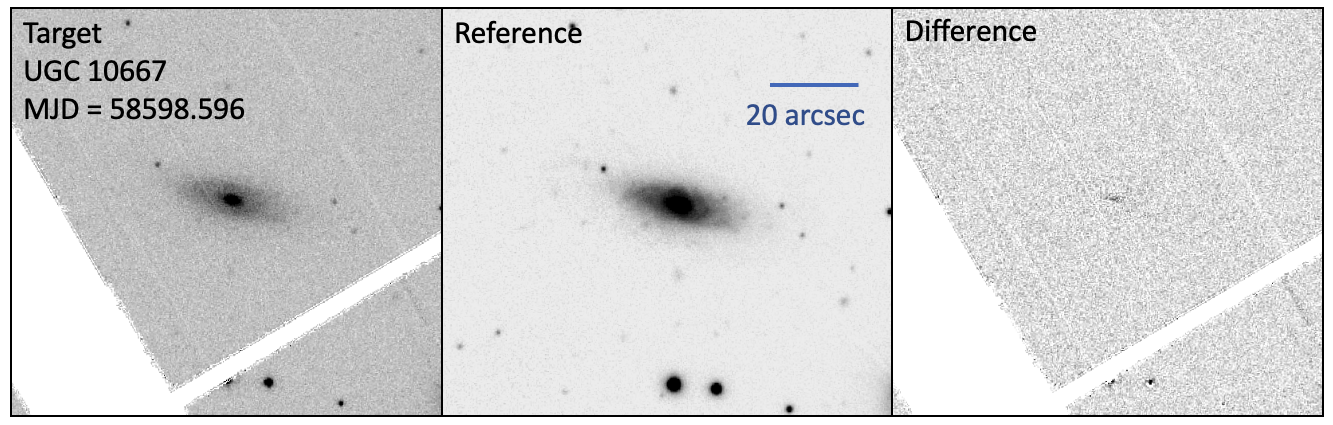}
    \caption{The Pan-STARRS1 images and difference images of the galaxy \HostGal\ from MJD = 58598.5957604. This  \ips-band image was taken 6.0\,hrs after GW190425 merger time and 3.5\,hrs hrs after FRB190425. As discussed in Section\,\ref{sec:PS1}, the excess flux at the core of the galaxy is either a difference image residual or low level AGN activity and there is no evidence of a transient source to typical depths of \ips$>21.6$ (north up and east left).}
    \label{fig:UGC10667_PS1}
\end{figure*}

\section{Observations and data}

\subsection{Pan-STARRS1 observations and data}
\label{sec:PS1}
The Pan-STARRS (PS) is a dual 1.8\,m telescope system (Pan-STARRS1 and Pan-STARRS2) each equipped with a 1.4 Gigapixel camera  located at the summit of Haleakala on the Hawaiian island of Maui \citep{2016arXiv161205560C}. 
The data for this paper were all taken with the Pan-STARRS1 (PS1) 
telescope and camera.  The  
0\farcs26 arcsec pixels give a focal plane of 3.0 degree diameter, that corresponds to a field-of-view area of 7.06 sq deg. It is equipped with a filter system, denoted as \grizy\ as described in  \citet{2012ApJ...745...42T}.
The Pan-STARRS1 Science Consortium
3$\pi$ Survey produced  $grizy_{P1}$ images of the whole sky north of $\delta = -30\degree$
\citep{2016arXiv161205560C}. 
We also have proprietary $i_{P1}$ data between $-40\degree < \delta < -30\degree$. These data provide reference images for immediate sky subtraction. Images from Pan-STARRS1 are processed immediately with the Image Processing Pipeline \citep{2020ApJS..251....3M,2020ApJS..251....4W}.

The individual exposure 
frames (called warps) are astrometrically and photometrically calibrated \citep{2020ApJS..251....6M}. 
The 60 CCDs in GPC1 are processed individually and warped onto a fixed tessellation of skycells as described in  \cite{2016arXiv161205560C}, each of which is typically 24$\times$24\arcmin.  
Overlapping exposures can be co-added together (on the skycell tessellation) 
with median clipping applied (to produce nightly stacks). The Pan-STARRS1 3$\pi$ reference sky images are subtracted from both the warps and the stacks 
\citep{2020ApJS..251....4W} and photometry carried out on the resulting difference image \citep{2020ApJS..251....5M}. These individual detections are ingested into the Pan-STARRS Transient Server database at Queen's University Belfast and assimilated into distinct objects with a time variable history. A series of quality filters are applied using the IPP image attributes and known asteroids and variable stars are removed. The objects remaining are cross-matched with all catalogued galaxies, AGN, CVs and historical transients \citep{2016MNRAS.462.4094S} and simultaneously a machine learning
algorithm is applied to image pixel stamps at each transient position \citep{2015MNRAS.449..451W}. This reduces the bogus detections to a manageable number for human scanning. During the first three LIGO-Virgo observing 
runs we had a programme in place  to cover the LIGO-Virgo 
skymaps for optical/near-infrared counterpart 
searches \citep{2016MNRAS.462.4094S,2016ApJ...827L..40S} and can typically cover 500-1000 square degrees
per night multiple times with one Pan-STARRS telescope. At the time of \GWsource, we 
were using Pan-STARRS1 (PS1) as the primary search facility for optical counterparts to 
GW sources. 

\GWsource\ was discovered at 
MJD=\MJDGW\ or \UTGW\
\citep[data from][]{GW190425LVSCDataset}, 
and announced publicly with an initial localisation skymap 
in a Gamma-ray coordination network (GCN) circular 95 minutes later at 
MJD=\alertMJDGW\
\citep{GCN190525z}. 
We began observing the initial LIGO-Virgo \initialMap\ skymap \citep{2016PhRvD..93b4013S}
with PS1 beginning at 
MJD=\MJDPSo, or
\MJDdiffoPS\, 
hours after the BNS merger time.  
A series of dithered and overlapping 45\,sec 
exposures were taken in the $i$-band over a period of 2hrs. 
The images were typically taken in a set of $4\times45$\,sec, separated across 
1hr or so to identify and remove moving objects. The images can either be processed 
individually or co-added to create a single stacked image for each Pan-STARRS skycell \citep{2016arXiv161205560C}. Standard processing as described above was immediately 
carried out on all the individual 45\,sec exposurs and these had
typical limiting magnitudes of 
\ips$> 21.3\pm0.3$\,AB. 
These values represent the median of the 3.5$\sigma$ upper limits of the processed skycells and the standard deviation of the sample.
While we used the \initialMap\ skymap to define our pointings at the time of the event, an updated map is available from the GWTC-2 release \citep{2021PhRvX..11b1053A} and all probability sky coverage in this paper is with respect to that map
(\texttt{gw190425z$\_$skymap.multiorder.fits})\footnote{\texttt{gw190425z$\_$skymap.multiorder.fits}  is available on  \texttt{https://gracedb.ligo.org/superevents/S190425z/view/}}.

On this first night we covered 1266 square degrees of the GWTC-2 skymap,  which 
corresponds to a cumulative probability coverage of 23.8 per cent. This was incremented over the first 3 days to cover 1374 square degrees (24.9 per cent). 
At the time, we released twenty five transients detected in the skymap \citep{2019GCN.24210....1S,2019GCN.24262....1S} 
but none of these emerged as a compelling candidate for an optical counterpart of  \GWsource\ lying in a host galaxy within 
the GW constrained redshift range. A number of 
candidates were followed up, and PS19qp showed a red continuum \citep{2019GCN.24217....1N,2019GCN.24221....1J} but it was
subsequently classified as a type Ic supernova suffering significant extinction  \citep{2019GCN.24230....1M,2019GCN.24358....1D,2019GCN.24295....1M}.
Figure\,\ref{fig:PS1coverage} shows the Pan-STARRS1 skymap coverage over the first three days. 

In the context of FRB190425 and its most likely host galaxy \HostGal\ 
(RA=\HOSTCoordRA, DEC=\HOSTCoordDEC),
we covered the position of this galaxy with PS1 within a few hours of both the FRB and the GW signals.  A single 45\,sec 
\ips-band image was taken at 58598.5957604, which is \MJDdiffgal\ after GW190425 merger time.
The image is of good quality, taken at an airmass of 1.04 and with image FWHM of 0.97\arcsec.
The image is shown in Figure\,\ref{fig:UGC10667_PS1}
 along with the PS1 3$\pi$ reference image \citep{2016arXiv161205560C} and the difference image created from subtracting the latter from the target image. There is no transient source
 visible in the PS1 images either in or around \HostGal.  There is a residual at the core of the galaxy which is almost certainly an image subtraction artefact and this is visible in historic monitoring of this sky region. 
 We estimate a 3.5$\sigma$ limit  of in the skycell of the image containing \HostGal\ to be $i > 21.6$. 
The limit is estimated using the method described in \cite{2020ApJS..251....5M}, in which the 
 the flux and variance images are smoothed with a  circularly symmetric Gaussian kernel and
 a significance image in signal-to-noise units is generated.  
A manual check of the background sky noise in a PSF aperture and locating the faintest sources deteted and visible in the image yields a detection limit of between \ips\ $>21.5\,{\rm to}\,21.8$, corroborating the PS1 processing method.   The circular radius around the 
 core of \HostGal\ which contains good and complete pixel data is 32\farcs0, or a projected 
 galactocentric distance in the sky plane of the galaxy of $R_{\rm g}=20.5$\,kpc. 
 Therefore we can say with reasonable confidence that there is no optical transient within 
 20.5\,kpc of \HostGal\ to a limiting mag of \ips$> 21.6$, at 6hrs after \GWsource\ merger
 time.  We revisited this sky region +13.2\,days and +35.2 days later during routine 
 PS1 sky survey operations. A quad of images was taken on each occasion ($4\times45$\,sec), 
 in the \wps\ filter and these were combined to create a nightly stack (a 180\,sec 
 exposure). No transient source was detected in the difference images to \wps$>23.5$ and 
 \wps$>22.6$ on either night respectively. A summary of the limiting magnitudes is listed in Table\,\ref{tab:obslimits}.
Although a projected offset of 20\,kpc would enclose most short GRBs 
\citep{2022ApJ...940...56F} and the few candidate kilonovae known, the recent GRB230307A and its associated kilonova was observed at an offset of 40\,kpc from its likely host \citep{2023arXiv230702098L,2023arXiv230800638Y,2023arXiv230800633G}. The ATLAS data described in Section\,\ref{sec:atlas} do not suffer from this pixel chip gap  issue.

\cite{FRB190425host} highlight six other galaxies that they estimate had a non-zero 
probability of being the candidate host of \FRBsource, in their methodology. 
The probabilities of any of them being the host ranged from 1 to 3.3  per cent
 \citep[Table 1 in][]{FRB190425host}. We covered all apart from WISEAJ170930.73+213633.8 with PS1 imaging on the first night and no positive and significant transient sources were detected in the difference images. No automated detections were found and all images were inspected visually. WISEAJ170930.73+213633.8 (probability of being the host of \FRBsource, 
 $P_{\rm PATH}=0.0311$) fell on a chip gap and no definitive conclusion on transient emission can be drawn. 

\subsection{ATLAS observations and data}
\label{sec:atlas}
At the time of writing, ATLAS is operating as a four telescope survey system with 
identical units in Haleakala and Mauna Loa (in Hawaii), El Sauce (Chile) and Sutherland 
(South Africa). However during the O3 observing run, the  two operational
telescopes were the northern units. As described in \cite{2018PASP..130f4505T}, 
each ATLAS unit is a "Wright Schmidt" type telescope with a 0.65m primary and 
a Schmidt corrector providing a 0.5m clear aperture. The detectors are STA-1600 CCDs, 
which are arrays of 10560$\times$10560 9$\mu$m pixels. The pixel
scale of 1\farcs89 gives a field-of-view of 28.9 square degrees for each camera. 
In normal survey mode in 2019 we were typically covering the sky north of $\delta >-45^{\circ}$ 
every 2 nights. During the O2 and O3 observing runs we frequently adjusted the 
ATLAS survey schedule to promptly cover GW maps, with no loss to the primary near-Earth asteroid mission.
We discovered the fast transient 
ATLAS17aeu which turned out to be a GRB afterglow, coincidentally in the skymap of the binary black hole merger
GW170104 \citep{2017ApJ...850..149S}

After the GW190425 alert, we scheduled sequences of 30 sec images in the ATLAS $o$-band, and at each pointing position a sequence of quads (4$\times$30 sec) was taken.  A summary of our observations was posted  by  \cite{2019GCN.24197....1M}. 
The images were processed with the ATLAS pipeline and reference images subtracted from each one
\cite{2018PASP..130f4505T}.  Transient candidates were run through our standard filtering procedures within the ATLAS Transient Science server \citep{2020PASP..132h5002S}. After quality control filters and real-bogus labelling with machine learning algorithms,  candidates were
spatially cross-matched with known minor planets, and star, galaxy,
AGN and multi-wavelength catalogues.  We began observing the northern part of the skymap within the first
hour of the preliminary notice.  During the first night, ATLAS covered 2799 sq. degrees of the 90 per cent credible region of the  GWTC-2 skymap
and covered a sky region totalling of 32.7 per cent of the probability area. By the third night of observing this was incremented to 4560 sq degrees and 41.6 per cent. The ATLAS coverage of the skymap is presented in Fig.\,\ref{fig:ATLAScoverage} and the 5$\sigma$ limiting magnitudes of the individual exposures are illustrated in Fig.\,\ref{fig:limits}. The median and standard deviation of the limiting magnitudes are $o > 19.2 \pm 0.3$. 

In \cite{2019GCN.24197....1M}, 
we flagged 25 transients but all were either already known, or we detected
previous flux in our own forced photometry in images before the merger.  No further convincing counterpart candidates were found brighter than $o\sim19.5$  which were plausibly associated with a galaxy within 100-200 Mpc (i.e. less than 50 kpc separation).  We reported 5 marginal candidates and noted they required independent confirmation \citep{2019GCN.24197....1M}, but all five were not recovered by other surveys and therefore were likely noise artefacts \citep[e.g.][]{2019GCN.24211....1N}. 

In the context of FRB190425 and its most likely host galaxy \HostGal, 
we covered the position of this galaxy with ATLAS within 6.9 hours after the GW signal.  A single 30\,sec exposure covered the coordinates of \HostGal\
(the quad was not completed at this sky position) and no transient flux is observed 
to a limiting magnitude of $o>18.2$ (this was one of the poorer images on the night of 58598). 

ATLAS also covered this position at +1.23 days after 
GW190425 and this time the $4\times30$ $o$-quad was completed in good conditions. 
The 4 separate difference 
image frames were co-added and no transient is visible within several arc minutes of \HostGal\ to a 3.5$\sigma$ limiting magnitude of $o>20.6$. A summary of epoch and observations is given in Table\,\ref{tab:obslimits}. 

\subsection{Zwicky Transient Facility observations and public data of \HostGal}
\label{sec:ztf}
The Zwicky Transient Facility (ZTF) observed the skymap in some of its public survey mode as described in \citep{2019ApJ...885L..19C}. No transient source is found within 30 arcseconds of \HostGal\ in the public stream ingested by the Lasair broker\footnote{https://lasair-ztf.lsst.ac.uk}  \citep{2019RNAAS...3...26S}. 
The ZTF survey \citep{2019PASP..131a8002B} allows forced photometry to be run at any position in the public data and the images to be requested \citep{2023arXiv230516279M}. 
We forced photometry at five positions at and around
\HostGal\ on 3 nights after  the time of \GWsource\ (58598.414 to 58600.393) and inspected the difference images. No source was detected apart from what appears to be a difference image residual at the core of \HostGal\ on  58599.397, similar to the PS1 residual in Figure\,\ref{fig:UGC10667_PS1}. The 3.5$\sigma$ limits are listed in Table\,\ref{tab:obslimits}.

\begin{table}
	\centering
	\caption{The 3.5$\sigma$ limits of the Pan-STARRS1, ATLAS  and ZTF images of \HostGal\ around the time of GW190425. The epoch refers to the time of the image compared to the merger time of GW190425, in units of days.}
	\label{tab:obslimits}
	\begin{tabular}{llllll} 
	\hline
	MJD & Epoch & Telescope & ExpTime & Filter & limit \\
	\hline
    58598.414317 & +0.06775  & ZTF   & 30s  & $g$   & $>20.6$\\   
    58598.445932 & +0.09936  & ZTF   & 60s  & $r$   & $>21.0$\\   
    58598.595760 & +0.24919  & PS1   & 45s  & \ips\ & $>21.6$\\
    58598.635555 & +0.28899  & ATLAS & 30s  & $o$   & $>18.2$\\ 
    58599.254896 & +0.90833  & ZTF   & 90s  & $g$   & $>21.3$\\ 
    58599.396968 & +1.05040  & ZTF   & 90s  & $r$   & $>20.4$\\
    58599.576363 & +1.22979  & ATLAS & 120s & $o$   & $>20.6$\\  
    58600.392824 & +2.04625  & ZTF   & 30s  & $g$   & $>21.1$\\     
    58611.567044 & +13.22047 & PS1   & 180s & \wps\ & $>23.4$\\
    58633.524370 & +35.17780 & PS1   & 180s & \wps\ & $>22.6$\\

		\hline
	\end{tabular}
\end{table}

\begin{figure}
	\includegraphics[width=\columnwidth]{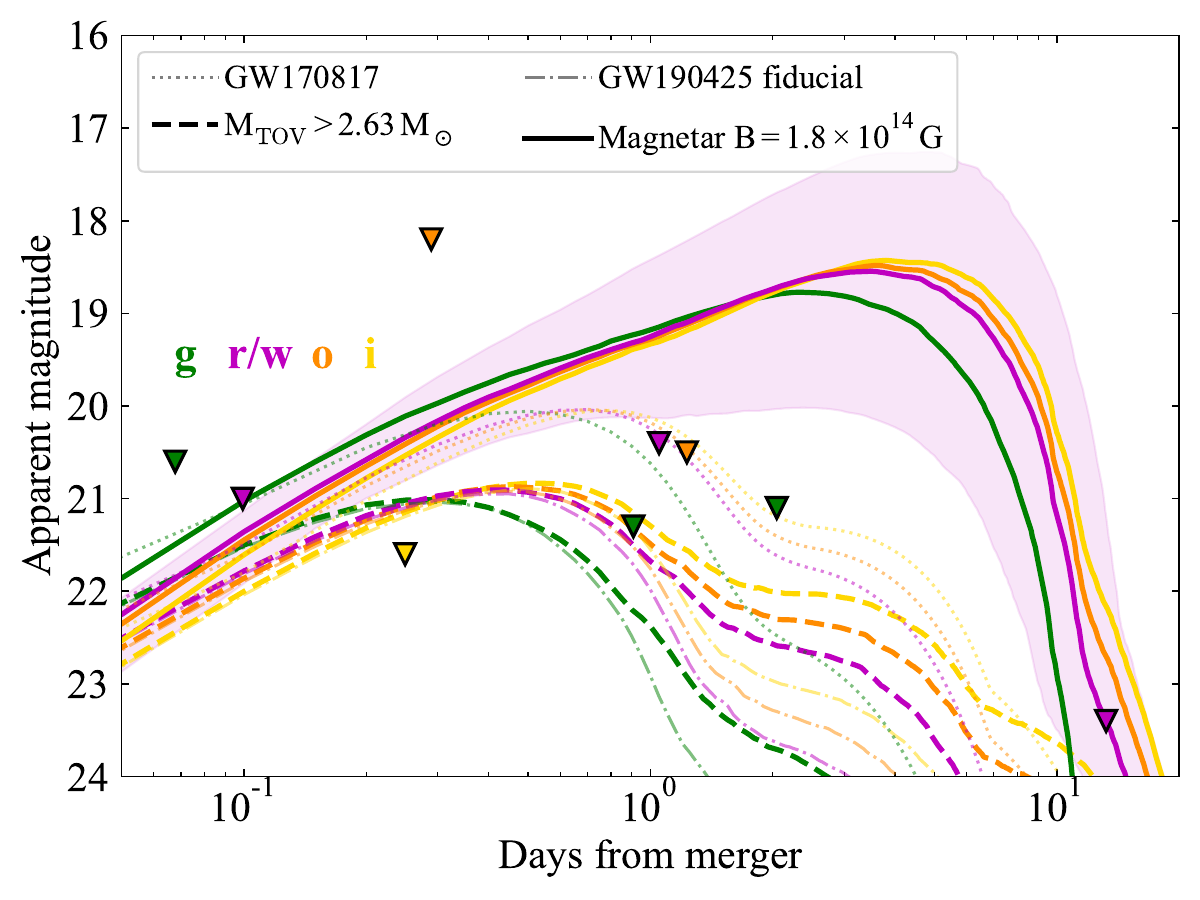}
    \caption{The upper limits measured by PS1, ATLAS and ZTF in images taken of \HostGal\ are plotted as inverted triangles. The  \wps\ filter is very close to $r$-band and are plotted in the same colour. Three models 
    of kilonova emission are plotted. The dotted and dashed lines are radioactively powered kilonova emission of GW170817 (AT2017gfo) and the fiducial model of ejecta mass of GW190425 estimated by \citet{Nicholl2021}, 
    see Section\,\ref{sec:KNUGC10667}.  A more luminous model with emission enhanced by a magnetar as described in Section\,\ref{sec:KNmagnetar}  is the solid line with model uncertainty regions. The models are calculated assuming a distance to \HostGal\ or \HostGal\ and the Milky Way foreground extinction as discussed in the text.
    }
    
    \label{fig:models}
\end{figure}

\section{Constraints on kilonova emission}
\label{sec:KNemission}

The general constraints on kilonova emission across the sky area covered jointly by Pan-STARRS and ATLAS are inconclusive, given that we covered  24.9 percent 
(\ips$> 21.3\pm0.3$)
and
41.2 percent ($o\gtrsim19.0$) integrated probability 
respectively. At the  estimated distance of GW190425 of $D_{\rm L}=$\GWdist\  \citep{GW190425LVC}, 
the PS1 data correspond to absolute magnitudes limits of $M_i \gtrsim -14.7^{-0.9}_{+1.3}$, assuming negligible extinction and combining the standard deviation of the limits with the distance uncertainty in quadrature. For ATLAS the obsevational constraints of ($o> 19.2 \pm0.3$) corresponds to absolute
magnitudes of  $M_o \gtrsim -16.8^{-0.9}_{+1.3}$. 

The Zwicky Transient Facility covered 46 per cent of the initial skymap and 21 per cent of the final skymap to magnitudes $g,r \sim21$ \citep{2019ApJ...885L..19C}. At the distance of $D_{\rm L}=$\HostDist\ there are plausible models of kilonova emission (calculated with varying ejecta masses and electron fractions) that would go undetected at the limits of Pan-STARRS, ZTF and ATLAS \citep[e.g.][]{2019MNRAS.489.5037B,Nicholl2021}. 

Other searches for counterparts were similar to, or less constraining than, the Pan-STARRS+ATLAS+ZTF combination in their coverage of the skymap 
\citep[e.g.][]{2019ApJ...880L...4H,2019ApJ...881L..26L,2020MNRAS.497.5518A,2020MNRAS.497..726G}.
There is little quantitative and meaningful limits that can be placed on the emission of a kilonova from this single event given the observing constraints. 

\subsection{Constraints on kilonova emission specifically in UGC 10667}
\label{sec:KNUGC10667}

We can directly and quantitatively assess the plausibility of optical emission from GW190425 if FRB190425 is associated with the GW emission and if \HostGal\ is the host galaxy as proposed by \cite{FRB190425} and \cite{FRB190425host}. 

To assess the significance of our non-detections of any optical emission from \HostGal, 
we compare to a range of representative kilonova light curve models generated using \textsc{mosfit} \citep{Guillochon2018,Villar2017,Nicholl2021}; these are shown in Figure \ref{fig:models}. Models are calculated in the ATLAS and Pan-STARRS filters and at the 
distance of \HostGal, adopting also the foreground reddening from NED. The simplest comparison is with the well-sampled, nearby kilonova AT2017gfo (from GW\,170817). For this we use the best fitting parameters from \citet{Nicholl2021} changing only the distance and extinction. The reader is  referred to  \citet{Nicholl2021} for details
of the model assumptions.  The Pan-STARRS \MJDdiffgal\ limit and the ATLAS limit at 
+1.22\,days both disfavour kilonova emission similar to that predicted for the GW170817 model (which matches the AT2017gfo data well). 

The GW190425 signal favoured a more massive merger than GW170817, and indeed more massive than any known Galactic NS binary, with a chirp mass $\mathcal{M}=1.44$\,M$_\odot$ (total mass $\approx3.4$\,M$_\odot$). This suggests a kilonova model calculated specifically for GW190425 may be more appropriate. We use a BNS-informed model from \citet{Nicholl2021}, with a narrow Gaussian prior on $\mathcal{M}$ and a flat prior on the mass ratio $0.8\leq q\leq1$. We also marginalise over uncertainties in the fraction of the remnant disk ejected ($0.1\leq \epsilon_{\rm disk}\leq0.5$), the fraction of lanthanide-poor ejecta from dynamical, rather than magnetic, processes ($0.5\leq\alpha\leq1$), and the fraction of polar ejecta heated by a gamma-ray burst jet ($0\leq\zeta_{\rm sh}\leq0.5$). The median light curve for GW190425 is $\approx0.7-1$ magnitude fainter at peak than GW170817 \citep{Nicholl2021}, though the uncertainties in parameters unconstrained by the GW signal result in a 90\% credible range spanning roughly $\pm1$ mag around the median.
The early PS1 data  point rules out the median model and excludes
$\approx 75\%$ of our model realisations. However, we note that since the models include several parameters without physically informed priors, this is not equivalent to ruling out a kilonova at 75\% confidence.

For these simplest GW170817 and GW190425 models, we have assumed a maximum stable NS mass $M_{\rm TOV}=2.17$\,M$_\odot$ \citep{Margalit2017,Nicholl2021}. However, the time delay between the GW and FRB signals favours a substantially larger $M_{\rm TOV}$ \citep{FRB190425}. A remnant NS in uniform rotation near break-up velocity is stable against collapse if its gravitational mass $M_{\rm rem}>1.2M_{\rm TOV}$. Thus to avoid prompt collapse of the rather massive remnant $M_{\rm rem}\approx 3.2$\,M$_\odot$ \citep{GW190425LVC}, an association between the GW and FRB signals requires $M_{\rm TOV}>2.6$\,M$_\odot$ \citep{FRB190425}. The authors also note that if the FRB results from the collapse of the remnant to a black hole (i.e., the remnant is not unstable indefinitely), we also have $M_{\rm TOV}<3.1$\,M$_\odot$. Marginalising over this uncertainty in $M_{\rm TOV}$, we find a light curve that looks essentially unchanged during the first $\sim 1$ day (compared to the fiducial set of models with $M_{\rm TOV}$ set at $2.17$\msol)
but is brighter by $1-2$ magnitudes during the next $\sim$ week. This is the phase when the intermediate opacity disk wind ejecta are expected to dominate the observed emission, and the increase in luminosity with $M_{\rm TOV}$ results from the more massive disk wind from a longer-lived remnant. These are plotted in Figure\,\ref{fig:models}, but the data we have are not constraining at the epochs that each set of models diverge. 

\subsection{Magnetar spin-down emission}
\label{sec:KNmagnetar}

If the link between GW190425 and FRB190425 were veracious and the physical picture is a supramassive (rotationally-supported) NS remnant subsequently collapsing into a black hole after 2.5 hours, the remnant must lose its rotational energy on this timescale. Merger remnants are expected to be rotating near break-up \citep{Radice2018}, with $P\simeq0.7$\,ms. For remnants with $M_{\rm TOV}<M_{\rm rem}<1.2M_{\rm TOV}$, the NS initially survives due to centrifugal support, and collapses once this is lost. Spin-down can occur through GW emission if the remnant has a quadrupole moment. However, we may expect that rotational energy loss is dominated by magnetic spin-down, particularly since the merger product is expected to have a strong magnetic field exceeding $10^{15}-10^{16}$\,G (\citealt{Price2006,Zrake2013,Kiuchi2023} -- and assuming the field is dominated by an ordered dipole; \citealt{DallOsso2009}). Spinning down through dipole emission on a timescale of 2.5 hours requires only a modest $B\sim{\rm  few}\times10^{14}$\,G \citep{FRB190425}. 

The rotational energy extracted from the remnant can greatly enhance the kilonova luminosity \citep[e.g.][]{Yu2013,Gao2015,Fong2021,Sarin2022}. \citet{Metzger2019} provide an analytic model for the luminosity resulting from dipole spin-down in a NS merger remnant, and show that it can boost the optical emission very significantly,  by up to  $\approx 4$ magnitudes. We refer the reader to that work for details, but in brief this model takes into account the typical dipole spin-down formula for the evolution of the spin period and magnetar luminosity often applied to supernova remnants \citep[e.g.][]{Kasen2010,2010ApJ...719L.204W}, and modifies it by a thermalisation efficiency (close to unity at early times) and the energy removed by electron-positron pair-creation at late times \citep{Metzger2014,Kasen2016}. The input luminosity goes to zero as soon as the remnant collapses. 

We have created a \textsc{mosfit} module to calculate magnetar-powered kilonova light curves using this framework. We assume a thermalisation efficiency $\epsilon_{\rm th}=1$,   an albedo of 0.5 for the pair cascade and a pair multiplicity of 0.1 \citep[as formulated in][]{Metzger2019}. The resulting energy injected by the magnetar as it spins down is converted to an output optical luminosity using the usual \citet{Arnett1982} model employed by \textsc{mosfit}. We have verified that this produces light curves in very good agreement with \cite{Metzger2019}. 
We fix the initial spin period at 0.7\,ms (i.e.~maximal spin), as expected from simulations. As the magnetar spins down, we compute the rotational energy at the time of collapse to a black hole following \citet{Margalit2017}. The energy available to power the transient is the difference between the initial rotational energy and that at collapse.

In the case of FRB190425, we fix $B=1.8\times10^{14}$\,G to give the appropriate time to collapse \citep{FRB190425}, resulting in no additional free parameters compared to the radioactive kilonova models. We marginalise over the chirp mass, mass ratio and ejecta parameters with the same priors as before.
This produces a luminous light curve, at all times brighter than the other models, and peaking later at $\sim 19$\,mag around 5-7 days after merger. The median model is strongly disfavoured by our Pan-STARRS and ATLAS observations of \HostGal. Furthermore,  in this case the credible range of the models does not overlap with our observational limits. 

We return to the quesition of what this can tell us about the plausibility of the GW190425 and FRB190425 connection. If the FRB-GW connection were true, \textit{and} if  the merger did occur in 
\HostGal\ then we can  exclude, with high confidence, that the merger 
produced a supramassive NS, spinning down by dipole emission on a timescale of hours. The working model to produce FRB190425 from a BNS merger, as proposed by \cite{FRB190425} is that of a supramassive neutron star that is highly magnetized \citep{2014ApJ...780L..21Z, 2014A&A...562A.137F}.  Hence the lack of detected optical emission disfavours but does not disprove, the FRB-GW link. Were such an FRB producing remnant formed, then we have shown it would likely have produced detectable optical emission. 
\cite{2023arXiv230600948B} propose that the FRB-GW association is unlikely as they find that a very low ejecta mass is required in order for the 400 MHz fluz to propogate through the material ejected in the merger and that the viewing angle requriements from the FRB and GW data are inconsistent. 


\section{Conclusions}
We promptly observed the LIGO-Virgo skymap of the BNS merger event GW190425 with Pan-STARRS and ATLAS  beginning several hours after the merger event. With Pan-STARRS1 we managed to cover a total integrated probability area of
24.9 per cent (to limiting magnitudes of \ips$> 21.3\pm0.3$)
over the first three days and with ATLAS we covered 41.2 per cent ($o>19.2\pm0.3$). 
These correspond to absolute magnitudes of $M_i \gtrsim -14.7^{-0.9}_{+1.3}$  and 
$M_o \gtrsim -16.8^{-0.9}_{+1.3}$. 
(assuming negligible extinction) with the errors dominated by the uncertainy in the distance to GW190425. The physical limits on 
an electromagnetic counterpart to GW190425 are not strong, given that approximately half the sky map was unobservable due
to solar conjunction - a problem that affected all wide-field searches for optical counterparts. However they do show the joint capability of the ATLAS and Pan-STARRS systems for gravitational wave followup, particularly as ATLAS is now a 4 unit system (and all-sky) and Pan-STARRS now is a twin facility on Haleakala. 

A recent proposed connection between GW190425 and the fast radio burst \FRBsource\ has emerged with a temporal  and spatial coincidence found by \cite{FRB190425}. If this association were to be physically true then it implies that a 
supramassive, rapidly rotating and magnetised neutron star was formed for at least a few hours after BNS merger (the GW and FRB signals were separated by 2.5hrs).
\FRBsource\ has been pinpointed to a most probable host galaxy, \HostGal\ which is at a compatible redshift with the 
distance to GW190425 \citep{FRB190425,FRB190425host}. With Pan-STARRS and ATLAS, we observed this host galaxy within 
a few hours of the FRB and GW signals. No optical emission was found.  We calculated samples of kilonova lightcurves with ejecta masses and radioactive heating based on 
the data from AT2017gfo and the physical parameters inferred from the GW data  of \GWsource\
\citep{Nicholl2021}. The Pan-STARRS limiting magnitude of $\ips>21.6$ at +0.25 days after GW190425 merger time, precludes an AT2017gfo type of kilonova and marginally disfavours a fiducial kilonova model based on the GW190425 data. 

The magnetised, rotating neutron star required to explain the FRB emission has a magnetic field of $B\sim1.8\times10^{14}$\,G which would result in an enhancement of the kilonova luminosity by magnetar powering. We calculate such models by fixing the magnetic field to that required by the FRB and marginalising over chirp mass, mass ratio and ejecta 
parameters. The rather luminous optical lightcurves are all ruled out by the limits from 
Pan-STARRS1 and ATLAS within the first +1.2\,days from merger. This excludes a supramassive NS, spinning down by dipole emission on a timescale of hours.  The lack of detected optical emission disfavours, but does not disprove, the FRB-GW link. If such an FRB-GW link were proven in the future \citep{FRB190425} then the FRB sky localisation and potential for immediate identification of a host galaxy \citep{FRB190425host} would be an extremely promising route to advance multi-messenger astronomy and further such coincidences should be searched for.

\section*{Acknowledgements}

Pan-STARRS is primarily funded to search for near earth asteroids through NASA grants NASA Grants NNX08AR22G,  NNX14AM74G. The Pan-STARRS science products for LIGO-Virgo follow-up made possible through the contributions of the University of Hawaii Institute for Astronomy and the Queen's University Belfast.
The Pan-STARRS1 Sky Surveys have been made possible through contributions by the Institute for Astronomy, the University of Hawaii, the Pan-STARRS Project Office, the Max Planck Society and its participating institutes, the Max Planck Institute for Astronomy, Heidelberg and the Max Planck Institute for Extraterrestrial Physics, Garching, The Johns Hopkins University, Durham University, the University of Edinburgh, the Queen's University Belfast, the Harvard-Smithsonian Center for Astrophysics, the Las Cumbres Observatory Global Telescope Network Incorporated, the National Central University of Taiwan, the Space Telescope Science Institute, and the National Aeronautics and Space Administration under Grant No. NNX08AR22G issued through the Planetary Science Division of the NASA Science Mission Directorate, the National Science Foundation Grant No. AST-1238877, the University of Maryland, Eotvos Lorand University (ELTE), and the Los Alamos National Laboratory.  

ATLAS is primarily funded to search for near earth asteroids through NASA grants NN12AR55G, 80NSSC18K0284,
and 80NSSC18K1575; byproducts of the NEO search include images and
catalogs from the survey area.  The ATLAS science products have been
made possible through the contributions of the University of Hawaii
Institute for Astronomy, the Queen's University Belfast, the Space
Telescope Science Institute, and the South African Astronomical Observatory.

Lasair is supported by the UKRI Science and Technology Facilities Council and is a collaboration between the University of Edinburgh (grant ST/N002512/1) and Queen’s University Belfast (grant ST/N002520/1) within the LSST:UK Science Consortium. ZTF is supported by National Science Foundation grant AST-1440341 and a collaboration including Caltech, IPAC, the Weizmann Institute for Science, the Oskar Klein Center at Stockholm University, the University of Maryland, the University of Washington, Deutsches Elektronen-Synchrotron and Humboldt University, Los Alamos National Laboratories, the TANGO Consortium of Taiwan, the University of Wisconsin at Milwaukee, and Lawrence Berkeley National Laboratories. Operations are conducted by COO, IPAC, and UW. This research has made use of ``Aladin sky atlas’‘ developed at CDS, Strasbourg Observatory, France. 

SJS, SS, SAS, LS, KS, DY acknowledge UKRI STFC grants ST/X006506/1, ST/T000198/1, ST/S006109/1 and 
ST/X001253/1. 
MN, SS, AA and XS are supported by the European Research Council (ERC) under the European Union’s Horizon 2020 research and innovation programme (grant agreement No.~948381). MN also acknowledges UK Space Agency Grant No.~ST/Y000692/1. TWC acknowledges the Yushan Young Fellow Program by the Ministry of Education, Taiwan for the financial support.
LJS acknowledges support by the European Research Council (ERC) under the European Union’s Horizon 2020 research and innovation program (ERC Advanced Grant KILONOVA No. 885281) and support by Deutsche Forschungsgemeinschaft (DFG, German Research Foundation) - Project-ID 279384907 - SFB 1245 and MA 4248/3-1.
This work was funded by ANID, Millennium Science Initiative, ICN12\_009 
CWS is grateful for support from Harvard University. 
This research has made use of the NASA/IPAC Extragalactic Database (NED), which is operated by the Jet Propulsion Laboratory, California Institute of Technology, under contract with the National Aeronautics and Space Administration.
For the purpose of open access, a Creative Commons Attribution (CC BY 4.0) licence will apply to any Author Accepted Manuscript version arising. 

\section*{Data Availability}
The data table for the ATLAS and Pan-SATRRS exposures (to produce Figures 1 and 2), the pixel images for Figure 3 will be made available when accepted either on the MNRAS website or on Research Data Oxford. The code for the models to produce Figure 5 is available on the MOSFiT https://github.com/guillochon/MOSFiT.



\bibliographystyle{mnras}
\bibliography{libsjs} 







\bsp	
\label{lastpage}
\end{document}